\documentclass[prl,aps,twocolumn,
superscriptaddress,floats,showpacs,10pt]{revtex4-1}
\usepackage{graphicx}
\usepackage{dcolumn}
\usepackage{bm}
\usepackage{amssymb}
\usepackage{amsmath}
\usepackage{textcomp}
\usepackage{color}

\begin{document}

\title{Stochastic Regimes in the Driven Oscillator with a Step-Like Nonlinearity}

\author{S. V. Bulanov}
\affiliation{Japan Atomic Energy Agency, Kansai Photon Science Institute, 
8-1-7 Umemidai, Kizugawa-shi, Kyoto, 619-0215 Japan}

\author{A. Yogo}
\affiliation{Institute of Laser Engineering, Osaka University, Suita 565-0871, Osaka, Japan}

\author{T. Zh. Esirkepov}
\affiliation{Japan Atomic Energy Agency, Kansai Photon Science Institute, 
8-1-7 Umemidai, Kizugawa-shi, Kyoto, 619-0215 Japan}

\author{J. K. Koga}
\affiliation{Japan Atomic Energy Agency, Kansai Photon Science Institute, 
8-1-7 Umemidai, Kizugawa-shi, Kyoto, 619-0215 Japan}

\author{S. S. Bulanov}
\affiliation{Lawrence Berkeley National Laboratory, Berkeley, California 94720, USA}

\author{K. Kondo}
\affiliation{Japan Atomic Energy Agency, Kansai Photon Science Institute, 
8-1-7 Umemidai, Kizugawa-shi, Kyoto, 619-0215 Japan}

\author{M. Kando}
\affiliation{Japan Atomic Energy Agency, Kansai Photon Science Institute, 
8-1-7 Umemidai, Kizugawa-shi, Kyoto, 619-0215 Japan}

\date{\today}

\begin{abstract}
A nonlinear oscillator with an abruptly inhomogeneous
restoring force driven by an uniform oscillating force exhibits stochastic properties under 
specific resonance conditions. This behaviour elucidates the elementary mechanism 
of the electron energization in the strong electromagnetic wave
interaction with thin targets.
\end{abstract}

\maketitle


\section{1. Introduction}

Chaos in nonlinear dynamics plays a fundamental role in understanding the properties 
of various physical systems \cite{TABOR}. A driven Duffing oscillator is an important example 
which demonstrates chaotic behaviour \cite{DNO1}. In particular, the problem of revealing the conditions 
for realization of chaotic regimes is of high importance 
for developing the theoretical description of strong electromagnetic wave interaction with 
matter. This is connected to the mechanisms of charged particle 
energization in collisionless plasmas (this is also called 'anomalous heating'), in their interaction 
with high intensity electromagnetic waves. As an example we note here the well 
known electron ``vacuum heating'' 
by laser radiation which occurs at the plasma-vacuum interface \cite{brunel}. Such processes have 
been invoked as an elementary mechanism of collisionless absorption of electromagnetic waves 
and of electron heating during  
high power laser interaction with various targets (e.g. see \cite{ThinFoil2}). Revealing of the 
elementary electron heating mechanism is important for developing laser plasma  sources 
of both fast electrons and ions.
 
Among the ion acceleration mechanisms  
 target normal sheath acceleration (TNSA) \cite{TNSAW} has been the most intensively investigated. 
The TNSA mechanism crucially depends on the plasma electron heating. 
It implies that the accelerated ion energy 
is proportional to the mean electron energy.  The anomalous heating 
of electrons at the plasma-vacuum interface has been noted in Refs. \cite{Taguchi, Aref},
where it was found that due to stochastic heating the energy of the electrons recirculating 
around a small size target can be substantially higher than 
the energy of the electron quivering in the field of the 
electromagnetic wave. Taguchi et al., \cite{Taguchi} in analysing 
the results of the particle-in-cell simulations 
have called this regime ``nonlinear resonance absorption''. 
Finding the conditions and criteria of the realization of 
such a regime 
 can have important consequences, in particular, for optimizing the laser-target parameters in developing 
a laser ion accelerator. Stochastic regimes in the electromagnetic wave 
interaction with a thin foil target have been found in theoretical papers \cite{ThinFoil0, KSI}. 
We note here also the electron heating by a strong electromagnetic field 
due to the electron rescattering off ions \cite{FMB-1999, ABWR-2003, BF-2006}. 

Below we analyze a model of a driven oscillator with a step-like 
nonlinearity in order to determine and elucidate the conditions 
when the stochastic vacuum heating takes place. The  3-dimensional 
model used bears the key features of the problem 
addressed in Refs. \cite{ThinFoil0, KSI, ZMS-2002, BPL-2007}, but in a significantly  
simpler mathematical approximation, 
because a full description of the 
charged particle interaction with electromagnetic waves implies 
a consideration of a 7-dimensional dynamical system.
In addition, such a mathematically simple  system can lead to a broader understanding of the underlying 
physical processes of electromagnetic field interaction with charged particles, thus leading to developing 
novel theoretical tools.

The paper is organized as it follows. In Section 2 we formulate 
the mathematical model of the nonlinear driven oscillator and analyze the properties 
of its free and driven oscillations. We obtain the discrete mapping and find the 
threshold of stochastic regimes. The trajectory diffusion is described by the 
Kolmogorov-Fokker-Planck equation (KFPE) whose analytical solution is presented. 
Using the KFPE solution we derive the asymtotic dependence on time of the average 
electron momentum. Then, in Section 3, we present the result of the 
numerical integration of the equations describing the trajectories of the 
driven oscillator with a step-like nonlinearity in nonrelativistic and 
relativistic limits, in regular and stochastic regimes. Here the time 
averaged electron kinetic energy is found numerically. At the end of the paper, 
in the Conclusion, we summarize the obtained results.

\section{2. Driven oscillator with a step-like nonlinearity}
\label{Sec2}

We consider the system of ordinary differential equations
\begin{equation}
\dot p+\varepsilon_p\, {\rm sign}{(x)}=a \cos t,
\label{eq:dotp}
 \end{equation}
\begin{equation}
\dot x=\frac{p}{\left(1+p^2\right)^{1/2}}.
\label{eq:dotx}
 \end{equation}
Here the sign function ${\rm sign}{(x)}=1$ for $x>0$ and ${\rm sign}{(x)}=-1$ for $x<0$, and the dots denote 
differentiation with respect to time, $t$. The driver force on the r. h. s. of Eq. (\ref{eq:dotp}) 
 represents the electromagnetic wave within the framework of the dipole approximation. 
 
 In the case of the electron circulating around a thin foil target, 
 $p$ and $x$ are the electron momentum and coordinate normalized by $m_e c$ and $c/\omega$, respectively, 
 where $m_e$ is the electron mass, $\omega$ is the driver field frequency, $c$ is the light speed in vacuum. 
 In Eqs. (\ref{eq:dotp}, \ref{eq:dotx}) the time variable is normalized by $\omega^{-1}$. The parameter
\begin{equation}
\varepsilon_p=\frac{2\pi n_0 e^2 l_0}{m_e \omega c},
\label{eq:eps_p}
 \end{equation}
introduced in Ref. \cite{VAV}, is proportional to the charge separation electric field, 
$E_{cs}=2\pi en_0 l_0$, normalized by $m_e\omega c/e$, for a thin foil with 
the charge surface density equal to  $en_0l_0$. The driver field normalized amplitude 
on the r. h. s. of Eq. (\ref{eq:dotp}) equals 
\begin{equation}
a=\frac{e E_0}{m_e \omega c}.
\label{eq:a}
 \end{equation}
If the driver field vanishes, $a=0$, the system of equations (\ref{eq:dotp}, \ref{eq:dotx}) describes 
free oscillations of a relativistic oscillator with an abruptly changing restoring force. 
The energy integral 
of Eqs. (\ref{eq:dotp}, \ref{eq:dotx}) is given by
\begin{equation}
\left(1+p^2\right)^{1/2}+\varepsilon_p |x|={\cal E}.
\label{eq:eng-int}
 \end{equation}
 Using this integral, we find the momentum and coordinate dependence in time for 
 the case of free oscillations:
\begin{equation}
p=p_m+ {\rm sign}{(x)} \varepsilon_p\, t, 
\label{eq:p(t)-free}
 \end{equation}
 \begin{equation}
x=x_0+ {\rm sign}{(x)}\frac{{\cal E}-\left[1+(p_m-\varepsilon_p\, t)^2\right]^{1/2}}{\varepsilon_p}, 
\label{eq:x(t)-free}
 \end{equation}
 with the constant ${\cal E}$  on the r. h. s. of Eq. (\ref{eq:x(t)-free}) equal to
\begin{equation}
{\cal E}=\left[1+(p_m-\varepsilon_p\, t]^2\right)^{1/2}+\varepsilon_p |x_0|.
\label{eq:eng-m}
 \end{equation}
 It is determined by the initial values of the momentum $p_m$ and coordinate $x_0$. 

 Using relationships (\ref{eq:p(t)-free}) and (\ref{eq:x(t)-free}) we 
find the period of nonlinear oscillations 
\begin{equation}
T=\frac{4 p_m}{\varepsilon_p}
\label{eq:T-free}
 \end{equation}
 depending on their amplitude $p_m$.
 
 Now we consider the case of driven oscillations, when $a \ne 0$, 
 i.e. the electron moves in the superposition of the 
 the static inhomogeneous electric field, $- \varepsilon_p {\rm sign}(x)$, 
 and oscillating homogeneous field, $a \cos t$. 
 This is described by the system of equations (\ref{eq:dotp},\ref{eq:dotx})
 with the
Hamiltonian
\begin{equation}
{\cal H}(x,p,t)=(1+p^2)^{1/2} + \varepsilon_p |x| - a x \cos t.
\end{equation}

The solution of Eqs. (\ref{eq:dotp},\ref{eq:dotx}) can be written in terms
 of piecewise-smooth functions.
Let the initial condition at $t=t_0$ be $x(t_0)=x_0$, and $p(t_0)=p_0$.
Let $t_k$ be consecutive zeros of the function $x(t)$, $k=1,2,3, \ldots$
(it is possible that $t_1=t_0$, but not required).

The solution to $p(t)$ in the interval $t\in [t_n, t_{n+1}]$ is
\begin{eqnarray}
p(t) = p_0 + \varepsilon_p S_0 
\left( 2\sum\limits_{k=1}^{n} (-1)^k t_k + t_0 + (-1)^{n+1} t \right)
\nonumber \\ 
+ a (\sin t-\sin t_0),
\end{eqnarray}
where
\begin{equation}
S_0 = \left\{
\begin{array}{ll}
{\rm sign}(x_0), & x_0\not=0, \\
{\rm sign}(p_0), & x_0=0, p_0\not=0, \\
{\rm sign}(\cos(t_0)), & x_0=p_0=0,\cos(t_0)\not=0, \\
-{\rm sign}(\sin(t_0)), & x_0=p_0=\cos(t_0)=0.
\end{array}
\right.
\end{equation}

Then for $x(t)$ we have
\begin{equation}
x(t)=\int\limits_{t_0}^{t} \frac{p(t)}{(1+p(t)^2)^{1/2}} dt .
\end{equation}

The consecutive zeros of the function $x(t)$ are determined by the implicit relationships
\begin{equation}
\int\limits_{t_k}^{t_{k+1}} \frac{p(t)}{(1+p(t)^2)^{1/2}} dt = 0,
\,\, 
k=0,1,2,\ldots\,.
\end{equation}

Loops are possible, if there exist integers $n\ge 2$ and $m\ge 1$, 
such that the momentum at $t_n$ is the same as at $t_1$, $p_n - p_1 = 0$,
and the phase of the oscillating component of the force is the same,
$t_n-t_1 = 2\pi m$.
This leads to the condition
\begin{equation}
2\sum\limits_{k=2}^{n} (-1)^k t_k - t_1 + (-1)^{n+1} t_n = 0 ,
\,\,
t_n = t_1 +2\pi m.
\end{equation}

 While the electron trajectory crosses the $x=0$ plane, 
 the sign of the static electric field changes abruptly. 
 This can be considered as a ``collision'', 
 during which the oscillating driver electric field produces work changing the 
 electron energy. 
 Depending on the phase, when the ``collision'' happens, the work can be either positive or negative, 
 resulting in the electron energy increasing or decreasing. 
 
 A characteristic electron momentum change during 
 the $x=0$ plane crossing can be estimated to be of the order of 
\begin{equation}
\Delta p=a.
\label{eq:Delta p}
 \end{equation}

 The enhancement of the energy transfer from the driver electric 
 field to the electron is expected to occur when 
 a resonant like condition takes place, i. e. when the periods of the motion 
 of the electron in subsequent periods 
 of the free oscillations with momenta $p$ and $p+\Delta p$ are co-measurable 
 with the period of the driver electric 
 field, $2\pi$. This yields
\begin{equation}
\frac{p}{\varepsilon_p}=\frac{\pi}{2}\,j
\label{eq:Tp}
 \end{equation}
 and
\begin{equation}
\frac{p+\Delta p}{\varepsilon_p}=\frac{\pi}{2}\,k
\label{eq:TpDp}
 \end{equation}
 with $j=1,2,3,...$ and $k=2,3,4,...\,$. From these expressions it follows that the resonance condition is
\begin{equation}
a=\frac{\pi (k-j)}{2}\varepsilon_p.
\label{eq:a-epsp}
 \end{equation}
This gives a relationship between the parameter $\varepsilon_p$ and 
the normalized driver amplitude, $a$, which 
can be rewritten in terms of the static and oscillating electric fields 
as $E_{cs}\approx E_0$. Here and further it 
is assumed that $k-j=1$. We note here that 
the condition (\ref{eq:a-epsp}) is equivalent to the criterion of thin slab relativistic transparency 
\cite{VAV, ThinFoil1, MIRR, REVFP} as well as determining the optimal parameters  
for ion acceleration by the 
laser light radiation pressure \cite{MIRR, REVFP, RPDA, SSB12}.

Using Eqs. (\ref{eq:Delta p}--\ref{eq:a-epsp}) we can find the electron momentum 
time dependence in the regimes of 
regular and stochastic acceleration. 

In the regular acceleration regime, the electron energy increases with  
$\Delta p=a$ during each half-period of free oscillations. Taking into account the free oscillation period 
dependence on the electron momentum (\ref{eq:T-free}) we obtain that the maximum of the momentum 
grows in proportion to 
the square root of time,
\begin{equation}
p_m=\left(a \varepsilon_p t/2\right)^{1/2}=a \left(t/\pi\right)^{1/2}.
\label{eq:p-treg}
 \end{equation}

\begin{figure}[]
  \begin{center}
    \includegraphics[height=6cm, width=7.1cm]{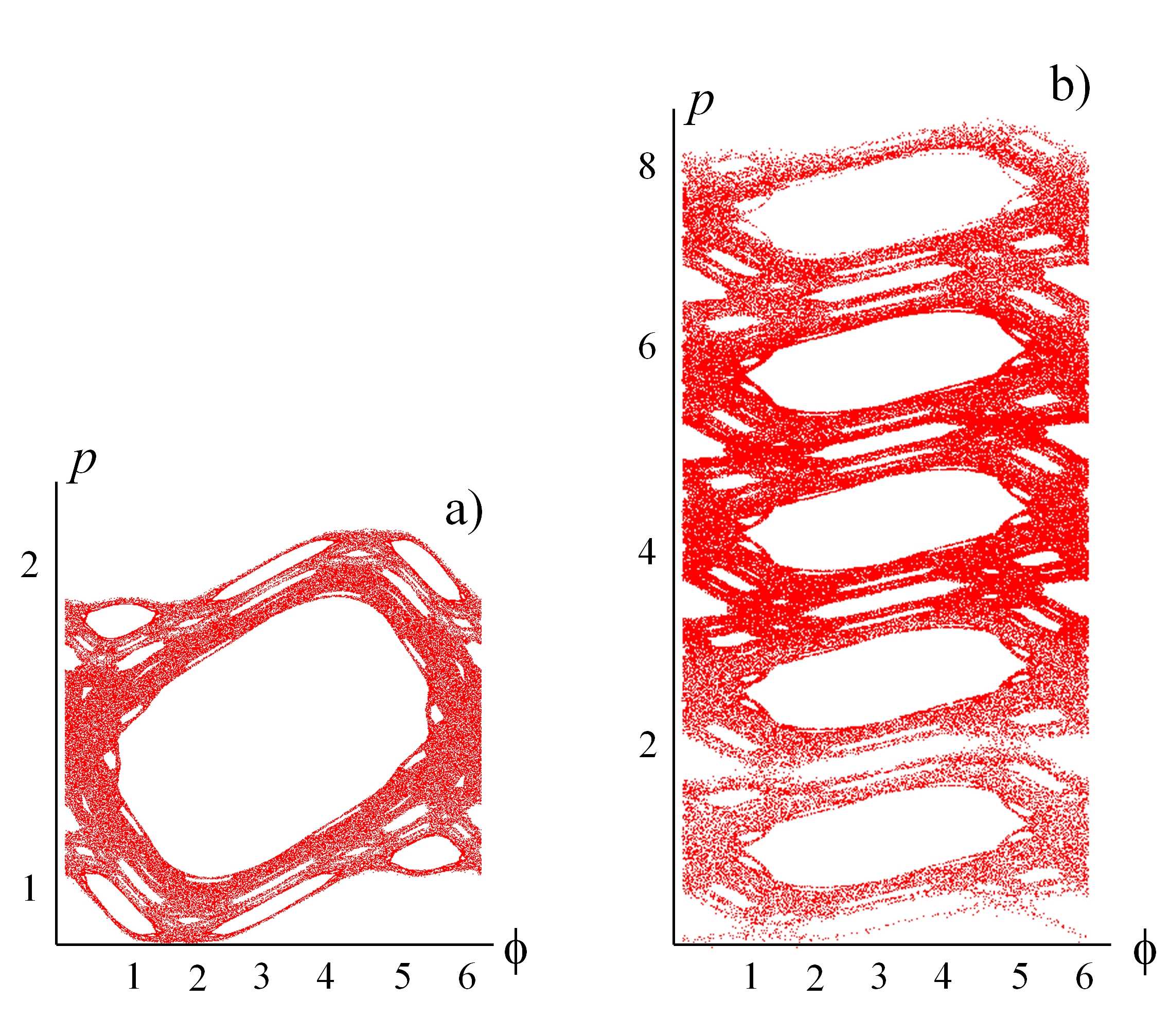}
  \end{center}
  \caption{Phase portrait $(\phi,p)$ of the mapping (\ref{eq:map-p},\ref{eq:map-phi}) for 
$\varepsilon_p=1$, $p_0=1$ and $\phi_0=\pi /3$.
a) The stochastic layer formation for $a=0.25$. 
b) The stochastic sea $a=0.3$.   
}
   \label{Fig1}
\end{figure}

In order to describe the stochastic acceleration let us assume that the trajectory crosses 
the $x=0$ plane at the instants of time equal to $t_n$. The momentum 
change is 
\begin{equation}
p_{n+1/2}=p_n-{\varepsilon_p}(t_{n+1/2}-t_{n})+a \sin (t_{n}),
\label{eq:map-p-p-n-n+1/2}
 \end{equation}
\begin{equation}
p_{n+1}=p_{n+1/2}+{\varepsilon_p}(t_{n+1}-t_{n+1/2})+a \sin (t_{n+1/2}).
\label{eq:map-n+1/2-n+1}
 \end{equation}
 In the limit of large momentum 
\begin{equation}
 |p_n|/a{\varepsilon_p}\gg 1
\label{eq:small-a}
 \end{equation}
 the time interval between two 
 subsequent crossings of the $x=0$ plane is equal to
\begin{equation}
t_{n+1}-t_n=\frac{4|p_{n+1}|}{\varepsilon_p}.
\label{eq:map-t-p-n-n+1}
 \end{equation}
 Here we have neglected the contribution of the order of $a$. Introducing the phase $\phi_n=t_n$ 
 we rewrite expressions  (\ref{eq:map-p-p-n-n+1/2}--\ref{eq:map-t-p-n-n+1}) in the form of the 
 Poncar\'{e} map:
\begin{equation}
p_{n+1}=p_{n}+2 a \sin (\phi_{n}),
\label{eq:map-p}
 \end{equation}
\begin{equation}
\phi_{n+1}=\phi_{n}+\frac{4|p_{n+1}|}{\varepsilon_p}\quad (\rm{mod}\,2\pi),
\label{eq:map-phi}
 \end{equation}
 which is also known as the standard map, corresponding to the model describing 
 stochastic Fermi acceleration \cite{Fermi1949, Fermi1954} 
 as a material point bouncing between two oscillating walls, e. g. see \cite{LL-1980, KK-2003}. 
 As is well known a Jacobian of the mapping is equal to unity,
\begin{equation}
\frac{\left(\partial p_{n+1}, \partial \phi_{n+1}\right)}{\left(\partial p_{n}, \partial \phi_{n}\right)}=1,
 \end{equation}
 i.e. the mapping conserves the phase volume. Due to the condition of the driver amplitude 
 smallness (\ref{eq:small-a}) the momentum change during subsequent crossing of the $x=0$ plane 
 is relatively small while the phase $\phi$ can change significantly. Using this fact we 
 can obtain from Eqs. (\ref{eq:map-p}) and (\ref{eq:map-phi}) that 
\begin{equation}
\phi_{n+1}\approx \phi_{n}+\frac{4|p_{n}|}{\varepsilon_p}+\frac{8|p_{n}|a}{\varepsilon_p}\sin(\phi_{n}).
\label{eq:map-phi1}
 \end{equation}
 This expression yields for the phase volume stretching along the $\phi$ coordinate 
\begin{equation}
|K|\approx \left|\frac{\delta \phi_{n+1}}{\delta \phi_{n}}-1 \right|\approx \frac{8 a|p_{n}|}{\varepsilon_p}\cos(\phi_{n}).
\label{eq:map-K}
 \end{equation}
 Since in the case of "highest probability" $|\cos(\phi_n)|\approx 1$, the criterion of the stochastic regime onset 
 is $|K|>1$ or
\begin{equation}
\frac{8 a|p_{n}|}{\varepsilon_p}>1.
\label{eq:map-K1}
 \end{equation}
In Fig. \ref{Fig1} we show the phase portraits $(\phi,p)$ 
of the mapping (\ref{eq:map-p},\ref{eq:map-phi})
 at the stochasticity threshold (a), and in the stochastic regime (b), when the particle energy can grow.

 Eq. (\ref{eq:map-p}) yields for the averaged squared deviation of the momentum change during one cycle

\begin{equation}
\overline{(\Delta p)^2}=\overline{(p_{n+1}-p_n)^2}=\overline{4 a^2 \sin^2(\phi_n)}=2{a^2}.
\label{eq:rms-p}
 \end{equation}
According to Eq. (\ref{eq:T-free}) the cycle duration is equal to $\tau=4 |p|/\varepsilon_p$. This gives
\begin{equation}
\overline{(\Delta p)^2}=\frac{a^2  \varepsilon_p\,\tau}{2 |p|},
\label{eq:rms-pt}
 \end{equation}
which shows that the momentum dependence on time 
has a character of diffusion with the diffusion coefficient 
\begin{equation}
D_{pp}=\frac{ \Delta p \Delta p}{\tau}=\frac{a^2 \varepsilon_p}{2 |p|}.
\label{eq:Dpp}
 \end{equation}

 Introducing the distribution function $f(t,p)$, which obeys the KFPE, 
 we write this equation as 
\begin{equation}
\partial_t f=\frac{1}{2} \partial_p \left(D_{pp} \partial_p f\right),
\label{eq:FP-Dpp}
 \end{equation}
 or
\begin{equation}
\partial_t f=\frac{a^2 \varepsilon_p}{4} \partial_p \left(\frac{1}{|p|}  \partial_p f\right).
\label{eq:FP-1/p}
 \end{equation}

 The solution of the KFPE for the initial 
 condition $f(t=0,p)=n_0 \delta(p)$, where $\delta(p)$ is the Dirac delta function, 
 can be easily found. It reads
\begin{equation}
f=n_0\frac{3^{1/3}}{(8 \varkappa t)^{1/3}\Gamma(1/3)} \exp \left( - \frac{|p|^3}{9\varkappa t}\right)
\label{eq:FP-f}
 \end{equation}
 where 
\begin{equation}
 \varkappa=\frac{a^2 \varepsilon_p}{2} 
 \end{equation}
 and $\Gamma(x)$ is the Euler gamma function \cite{GR}.
 
 Using Eq. (\ref{eq:FP-f}) we can find the time dependence of the average electron momentum, 
\begin{equation}
\overline{p}=\int p f(p) dp. 
\end{equation}
 It is given by
\begin{equation}
\displaystyle{\overline{p}=\frac{3^{2/3}\Gamma(2/3)}{2 \Gamma(1/3)}(\varkappa t)^{1/3}=
\frac{3^{2/3}\Gamma(2/3)}{2^{4/3} \Gamma(1/3)}}(a^2 \varepsilon_p t)^{1/3}.
\label{eq:p-tstoch}
 \end{equation}
 with the momentum dependence on time $p\sim t^{1/3}$.

The average kinetic energy,
\begin{equation}
 \frac{\overline{{\cal E}_e}}{m_e c^2}=\int_{-\infty}^{{+\infty}} [(1+p^2)^{1/2}-1]f(p)dp
 \end{equation}
 can be written in terms of the Meijer G - function \cite{GR} 
\begin{equation}
\frac{\overline{{\cal E}_e}}{m_e c^2}=-\left[1+\frac{G_{3,5}^{5,3}\left( \frac{1}
 {324(\varkappa t)^{2}}\left\vert \frac{1/6,1/2,5/6}{-1/3,0,0,1/3,1/2}\right. \right)}
 {48\times 3^{1/6}\pi^3(\varkappa t)^{1/3}}
  \right].
 \label{eq:p-tstochEn}
 \end{equation}
 In the limit of small but finite momentum this expression yields
\begin{equation}
 \frac{\overline{{\cal E}_e}}{m_e c^2}\approx \frac{\Gamma(1/3)\Gamma(1/6)\Gamma(5/6)}{4\pi^2}(\varkappa t)^{2/3}=
 1.6151 (\varkappa t)^{2/3}.
 \label{eq:p-tstocht0}
 \end{equation}
 When $t \to \infty$ we have
\begin{equation}
 \overline{\frac{{\cal E}_e}{m_e c^2}}\approx \frac{2^{2/3}\pi^{3/2}\Gamma(1/6)}{3^{4/3}\Gamma(1/3)^2\Gamma(2/3)^2}(\varkappa t)^{1/3}=
 1.051 (\varkappa t)^{1/3}.
 \label{eq:p-tstochtinfty}
 \end{equation}
The energy dependence on time, $\overline{{\cal E}_e}\sim t^{1/3}$, has been noted in Ref. \cite{ThinFoil0}.

The above derived standard map given by Eqs. (\ref{eq:map-p},\ref{eq:map-phi}), 
the KFPE equation (\ref{eq:FP-1/p}), 
and the KFPE solution (\ref{eq:FP-f}), showing the estimations 
for the stochasticity threshold (\ref{eq:map-K1}) and the heating rate (\ref{eq:p-tstoch}),
have been obtained by using certain assumptions, 
which correspond to the conditions imposed by Eqs. 
(\ref{eq:Delta p}, \ref{eq:small-a}, \ref{eq:map-phi1}). 
Since, as is well known, the properties and 
behaviour of dynamical systems in the stochastic regimes are very sensitive to the parameters, 
below we present the  results of numerical integration 
of the equations of motion in classical and relativistic limits. 
They provide additional information on 
the properties of a driven oscillator with a step-like nonlinearity.

\section{3. Results of numerical solutions of the equations describing driven oscillations}
\label{Sec3}
 
 In order to analyse in more detail the chaotic v.s. regular dynamics of the driven 
 nonlinear oscillator we have integrated Eqs. (\ref{eq:dotp}, \ref{eq:dotx}) numerically for different 
 parameters. In order to regularize the singularity on the left hand side of Eq. (\ref{eq:dotp}) we use 
 instead of ${\rm sign}(x)$ the function ${\rm Tanh}(x/l)$ with the width $l$ substantially smaller than
 the typical 
 particle displacement. In dimensional variables this requires 
 $l \ll eE/m_e \omega_{pe}^2$, where $\omega_{pe}=
 \left(4 \pi n e^2/m_e \right)^{1/2}$ is the Langmuir frequency. 
 
 The oscillator dynamics exhibits principally different behaviour depending 
 on the relative values of the parameter 
$\varepsilon_p$ and driver amplitude $a$. When the normalized driver 
amplitude is substantially less than unity, 
$a \ll 1$, we have the non-relativistic limit with nonlinear effects 
due to the abrupt dependence of the restoring force 
on the coordinate. In opposite limit, when $a \gg 1$, in addition there 
is a nonlinearity due to the relativistic 
relationship between the velocity and momentum (\ref{eq:dotx}). 

\subsection{Non-relativistic regime}

In the classical limit the behaviour of the driven nonlinear oscillations 
is determined by one dimensionless parameter,
$b=a/\varepsilon_p$, which can be written as 
\begin{equation}
b=\frac{E_0}{2\pi e n l}.
\label{eq:b-param}
 \end{equation}
This is equal to the ratio of the driver electric field to the electric 
field formed by a thin foil with the electric charge 
surface density $2\pi e n l$. The system of equations (\ref{eq:dotp}) and (\ref{eq:dotx}) takes the form
\begin{equation}
\ddot x+{\rm sign}\,(x)=b \cos (t).
\label{eq:ddotx-cl}
 \end{equation}
Here the $x$ variable is normalized by $c/\omega \varepsilon_p=m_e c^2/2\pi e^2 n l=d_e^2/l$ 
with $d_e=c/\omega_{pe}$ being the collisionless skin depth. We shall also use the notation $p=\dot x$.

In the case of free oscillations when the r. h. s. of Eq. (\ref{eq:ddotx-cl}) vanishes, 
i. e. $b=0$, the energy integral 
(\ref{eq:eng-int}) takes the form 
\begin{equation}
\frac {p^2}{2}+|x|={\cal E}
\label{eq:eng-int-cl}
 \end{equation}
 with the constant ${\cal E}={p_0^2}/{2}+|x_0|$. 
 The period of free oscillations is given by Eq. (\ref{eq:T-free}), which 
 in the notations used in this subsection is $T=4 p_m$. 

The resonance condition (\ref{eq:a-epsp}) in the non-relativistic limit yields
\begin{equation}
b=\pi/2.
\label{eq:a-epsp-cl}
 \end{equation}

 If the driver amplitude is substantially smaller than unity, for the initial conditions $p_0=0$ and $x_0=0$ 
 the trajectory is confined in the vicinity of the equilibrium $x=0$  within the region $|x| \ll l$.
 
\begin{figure}[]
  \begin{center}
    \includegraphics[height=6.3cm, width=8cm]{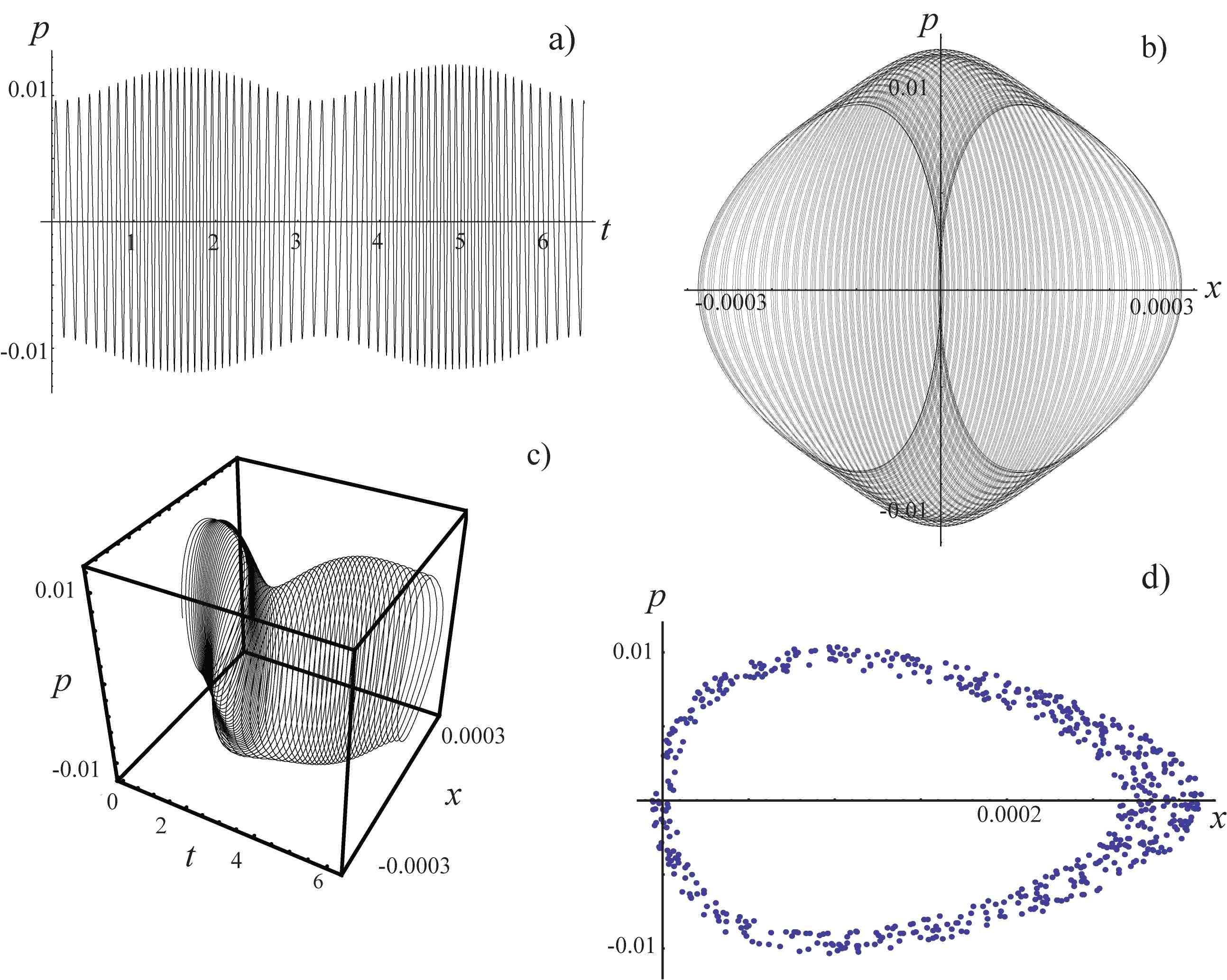}
  \end{center}
  \caption{Driven oscillations below the stochastic regime threshold 
  in the case of $b=0.775$, $l=0.000125$, $p_0=0$, and $x_0=0$. 
  a) The momentum $p$ v.s. time. 
  b) The phase plane $(x,p)$ for $t\in[0,10\times 2\pi]$.
  c) The configuration space $(t,x,p)$ for $t\in[0,10\times 2\pi]$.
  d) The Poincar\'{e} section showing the particle positions in the phase plane $(x,p)$ at the discrete time 
  with the time step 
equal to the period of the driving force, $2 \pi$.}
   \label{Fig2}
\end{figure}

 For the driver amplitude 
 approaching the threshold (\ref{eq:a-epsp-cl}) the oscillations become 
 nonlinear with the excursion exceeding the width $l$ as 
  seen in the case shown in Fig. \ref{Fig2} when the chosen parameters are close, 
 but below  the threshold (\ref{eq:a-epsp-cl}). The dependence of the momentum and coordinate on time 
 in Figs. \ref{Fig2} (a) and (c) correspond to nonlinear oscillations 
 with slowly varying period and amplitude. 
 In the phase plane, Fig. \ref{Fig2} (b), it is seen 
 that the trajectory is confined within finite domain. 
  In Fig. \ref{Fig2} (d) we plot the Poincar\'{e} section showing the particle positions 
 in the phase plane $(x,p)$ at the discrete time with the time step 
equal to the period of the driving force, $2 \pi$. 
The loop trajectory is slightly broadened due to approaching the stochastic regime.

There is a difference between the Poincar\'{e} sections presented here, and below 
in Figs. \ref{Fig3} (d), \ref{Fig4} (b,d,f), \ref{Fig5} (d), \ref{Fig6} (d), \ref{Fig8} (a), \ref{Fig11} (b),
 and in \ref{Fig12} (b), 
and those Poincar\'{e} sections which are shown 
in Figs. \ref{Fig1} (a) and (b). 
The driven oscillator dynamics is characterized by two time scales: the period of the driver force and 
 the time interval  between subsequent crossing by the trajectory the plane at $x=0$.
While in Fig. \ref{Fig1} and in Fig. \ref{Fig8} (b) each point in the plane $(\phi,p)$ 
corresponds to the particle momentum and the phase 
calculated for the time when the trajectory crosses of the plane at $x=0$, 
the points in the phase planes $(x,p)$ 
presented in Figs. \ref{Fig3} (d), \ref{Fig4} (b,d,f), \ref{Fig5} (d), \ref{Fig6} (d), 
\ref{Fig11} (b),  and in \ref{Fig12} (b) show 
the particle coordinate and momentum at the subsequent instants of time separated 
by the driver force period. 

For the driver amplitude slightly above the stochasticity threshold being equal 
to $b=0.795$ the oscillations become 
apparently stochastic as clearly seen in Fig. \ref{Fig3}. 
During the initial interval of time $t\in [0,50]$ 
the oscillations have a relatively low, gradually growing amplitude as seen in Figs. \ref{Fig3} (a)--(c). 
Then
it abruptly, during several periods, increases up to the level approximately 5 times 
higher than the driver amplitude.
The irregular oscillations can roughly be represented as a superposition of 
two modes with higher and lower amplitude.
There is also seen an intermittency when for some periods of time the oscillation 
amplitude becomes abruptly small and then 
increases again. The Poincar\'{e} section in Fig. \ref{Fig3} (d) demonstrates 
the trajectory broadening and appearance 
of islands as is typical for stochastic regimes \cite{TABOR}.

\begin{figure}[]
  \begin{center}
    \includegraphics[height=6.3cm, width=8.1cm]{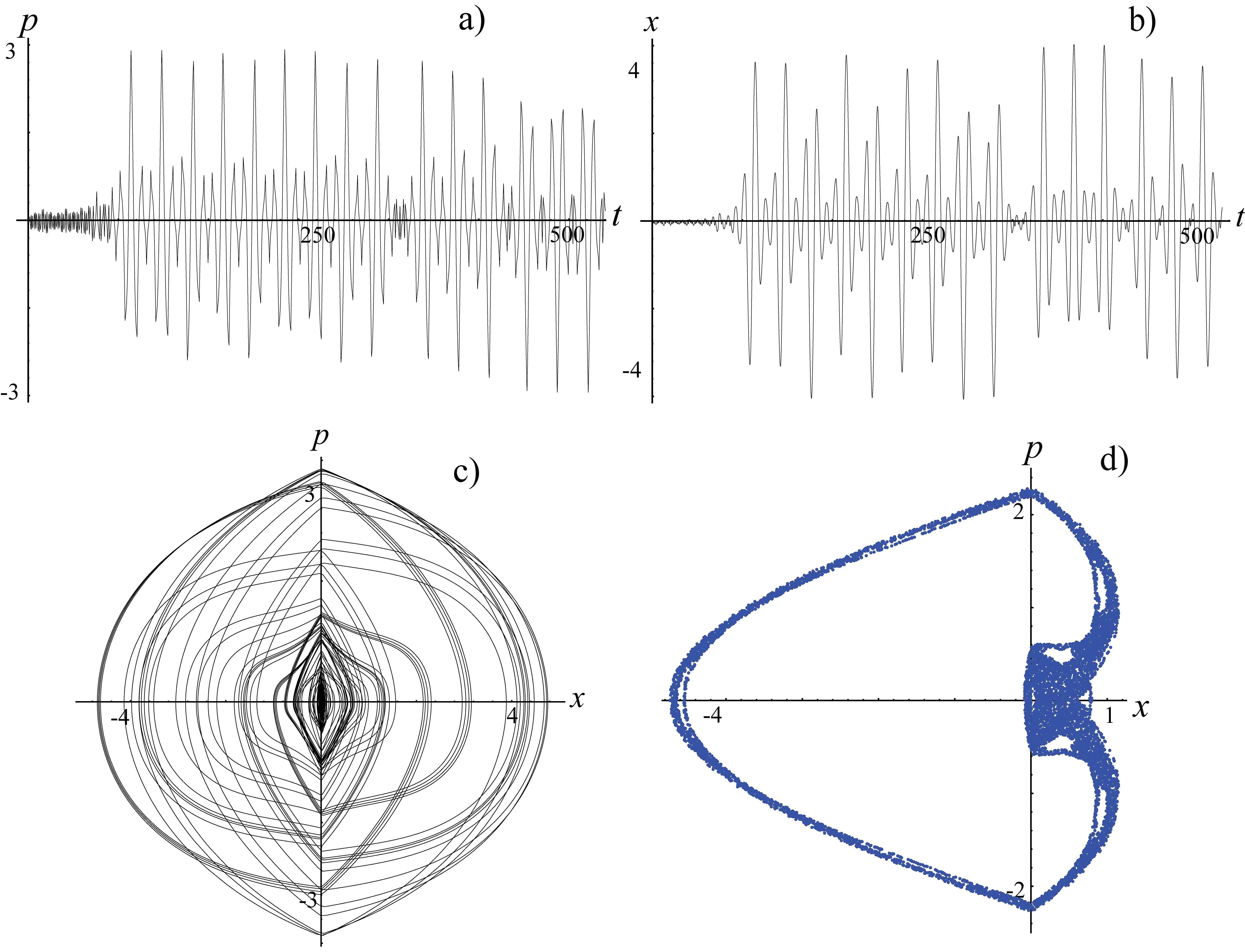}
  \end{center}
  \caption{Driven oscillations for the driver amplitude higher than the  
  stochasticity threshold for 
  $b=0.795$, $l=0.0125$, $p_0=0$, and $x_0=0$. 
  a)  The momentum $p$ v.s. time. 
  b) The coordinate $x$ v.s. time. 
  c) The phase plane $(x,p)$.
  d) The Poincar\'{e} section showing the particle positions 
  in the phase plane $(x,p)$ at the discrete time 
  with the time step 
equal to the period of the driving force, $2 \pi$.
  }   \label{Fig3}
\end{figure}

Further increase of the driver amplitude results in developed stochastic oscillations 
illustrated in Fig. \ref{Fig4}. If the driver amplitude is equal to 1, the corresponding phase plane and  
 Poincar\'{e} section (not shown here) are similar to ones presented in Fig. \ref{Fig3}. 
 If the driver amplitude, $b$, is equal to 5 the trajectory in the phase plane, Fig. \ref{Fig3} (a), 
 is confined in the gradually broadening stripes.
 When the driver amplitude, $b$, is equal to 9 the trajectory tightly fills the confinement 
 domain in the phase plane (see Fig. \ref{Fig4} (c)) showing ergodicity. The Poincar\'{e} sections in 
 Figs. \ref{Fig4} (b) and (d) demonstrate the formation of stability islands with the separatrices 
 and the regions with developed stochasticity. 
 The oscillation amplitude in Figs. \ref{Fig4} (a) and (b) remains 
 finite being of the order of the driver amplitude which is typical 
 for minimal chaos \cite{Chernikov, Dyson}.

\begin{figure}[]
  \begin{center}
    \includegraphics[height=8.25cm, width=8.cm]{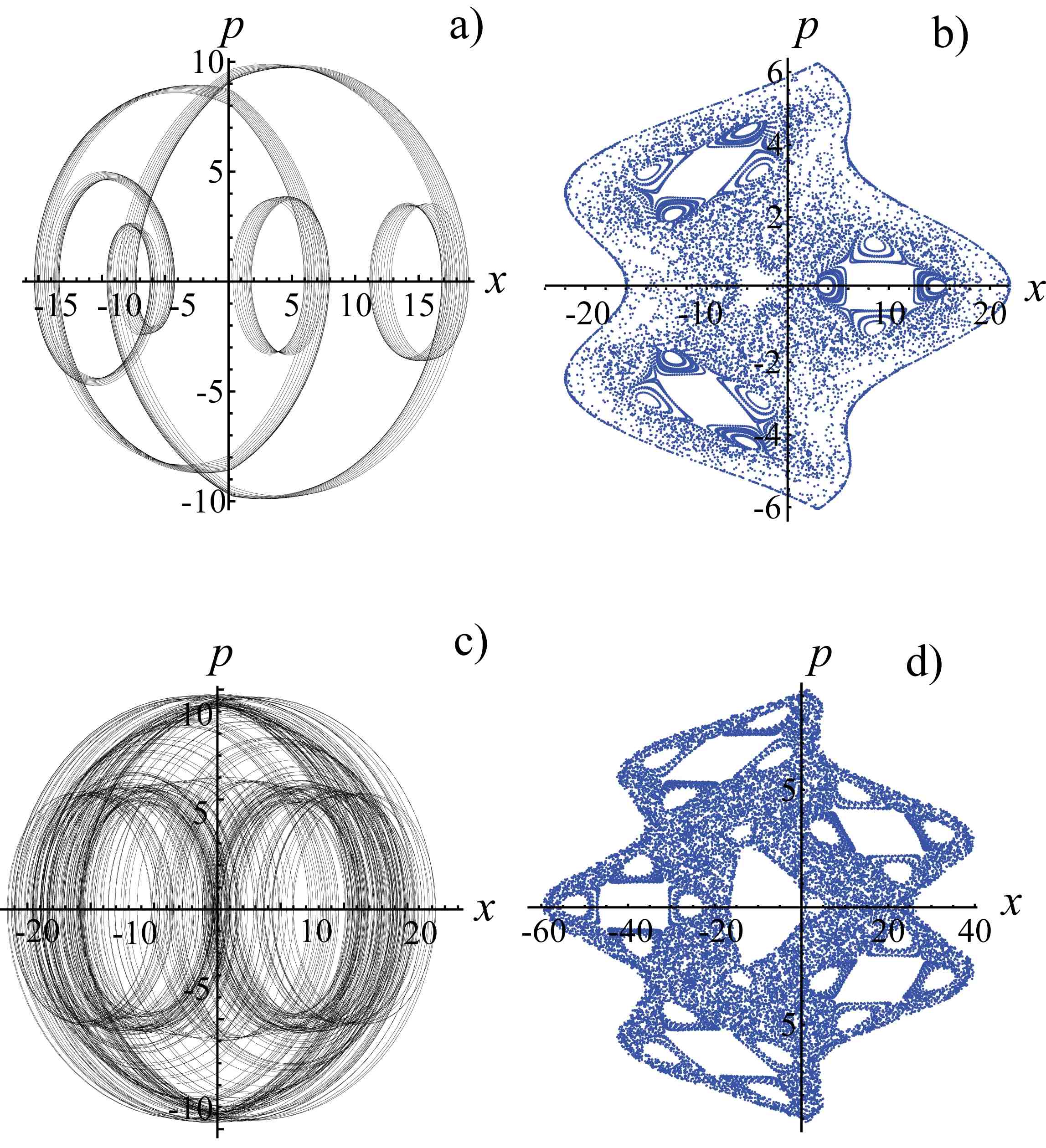}
  \end{center}
  \caption{Driven oscillations in  
  near-stochastic (a,b) 
  and stochastic (c,d) regimes for 
  $b=5$ and $9$, respectively, with $l=0.00125$. 
  a,c)  The phase planes $(x,p)$. 
  b,d) The Poincar\'{e} sections showing the particle positions 
  in the phase plane $(x,p)$ at the discrete time 
  with the time step 
equal to the period of the driving force, $2 \pi$.
  }   \label{Fig4}
\end{figure}

\subsection{Relativistic regime}

In the relativistic regime the behaviour of the system (\ref{eq:dotp}) and (\ref{eq:dotx}) 
is governed by two 
dimensionless parameters, $a$ and $\varepsilon_p$. In this limit, 
in addition to the nonlinearity due to the step like profile of the restoring force on the 
l.h.s. of Eq. (\ref{eq:dotp}),
 we have according to the 
special theory of relativity a nonlinear relationship between the particle velocity and momentum 
given by Eq. (\ref{eq:dotx}).

In the limit of small driver ampltitude 
$a\ll \varepsilon_p$ the trajectory is confined in the vicinity of an equilibrium $x=0$. 
The oscillation amplitude and frequency are of the order of $al/2 \varepsilon_p$ 
(in dimensional units it is $eE/m_e\omega_{pe}^2$) and $\left(2 \varepsilon_p/l \right)^{1/2}$,
 respectively. 
In dimensional units the oscillation frequency is equal to $\omega_{pe}$. The condition of the 
displacement smallness compared to the foil width $l$ is equivalent to the condition of 
low driver force frequency 
compared to the frequency of free oscillations. In turn it results in slow modulations of the oscillations.
 
\begin{figure}[]
  \begin{center}
    \includegraphics[height=6.3cm, width=8.1cm]{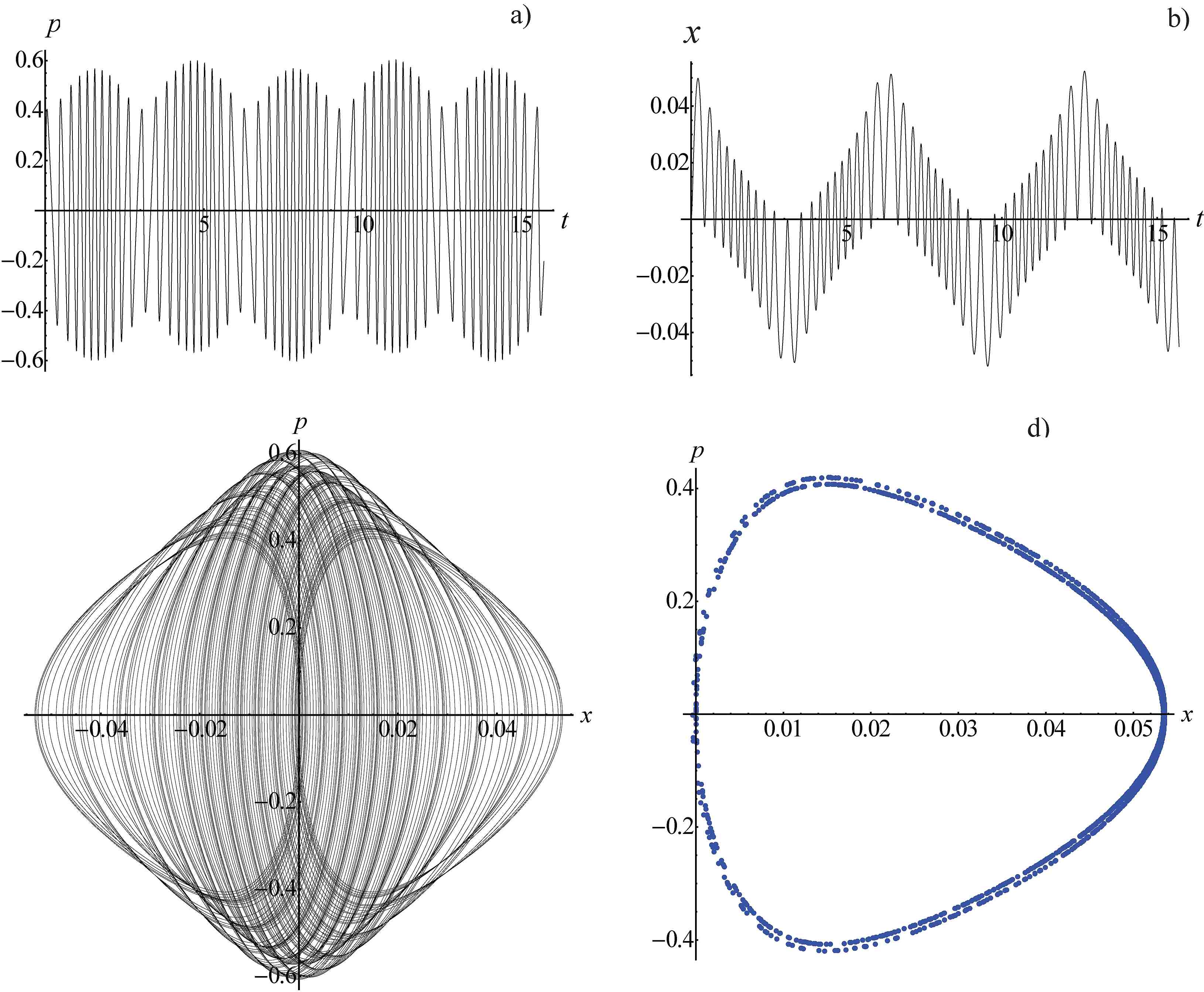}
  \end{center}
  \caption{Driven oscillations in nonresonant case for $\varepsilon_p=15.708$, 
  $a=0.52 (\pi \varepsilon_p/2)\approx 13.05$ and $l=0.0125$. 
  a) The momentum $p$ v.s. time. 
  b) The coordinate $x$ v.s. time. 
  c) The phase plane $(x,p)$ for $t\in[0,2\pi]$.
  d) The Poincar\'{e} section showing the particle positions 
  in the phase plane $(x,p)$ at discrete times, with the time step 
equal to the period of the driving force, $2 \pi$.}
   \label{Fig5}
\end{figure}

 Fig. \ref{Fig5} shows the case, when the driver amplitude being equal to $a=0.5 (\pi \varepsilon_p/2)$ 
 for $\varepsilon_p=15.708$ is close but lower than the resonant level.  Here and below we have
 choosen the width $l=0.05$, $a=15.298$, and $\varepsilon_p=15.708$. 
 For $\omega_{pe}/\omega=10$ this corresponds to the 
 foil width of the order of 10 nm.
 As we see, the momentum amplitude 
 $p_m\approx 0.1$ is relatively small (see Figs. \ref{Fig5} (a) and (b)). It remains small 
 for a significantly 
 long time interval.
 The corresponding oscillation period of the order of $4 p_m/\varepsilon_p\approx 3\times 10^{-2}$ 
 is substantially 
 shorter than the period of the driving force, which modulates the high frequency oscillations 
 as seen in Fig. \ref{Fig5} (a). The trajectory fills the region $(-0.003<x<0.003,-0.15<p<0.15)$ 
 of the phase plane shown in \ref{Fig5} (c) due to the anharmonicity of the nonlinear oscillator. 
 The driver force also modulates  the particle motion in the $x$ direction as 
 clearly seen in Fig. \ref{Fig5} (b). 
 The Poincar\'{e} section showing the particle positions on the phase plane $(x,p)$ 
 at the discrete time, with the time step 
equal to the period of the driving force, $2 \pi$, is plotted in Fig. \ref{Fig5} (d). 
We see that the trajectory is split into two sub-loops due to the oncet of the stochastic regime.
This corresponds to a superposition 
of the regular periodic circulation and the relatively weak stochastic motion.
 
\begin{figure}[]
  \begin{center}
    \includegraphics[height=10.4cm, width=8.1cm]{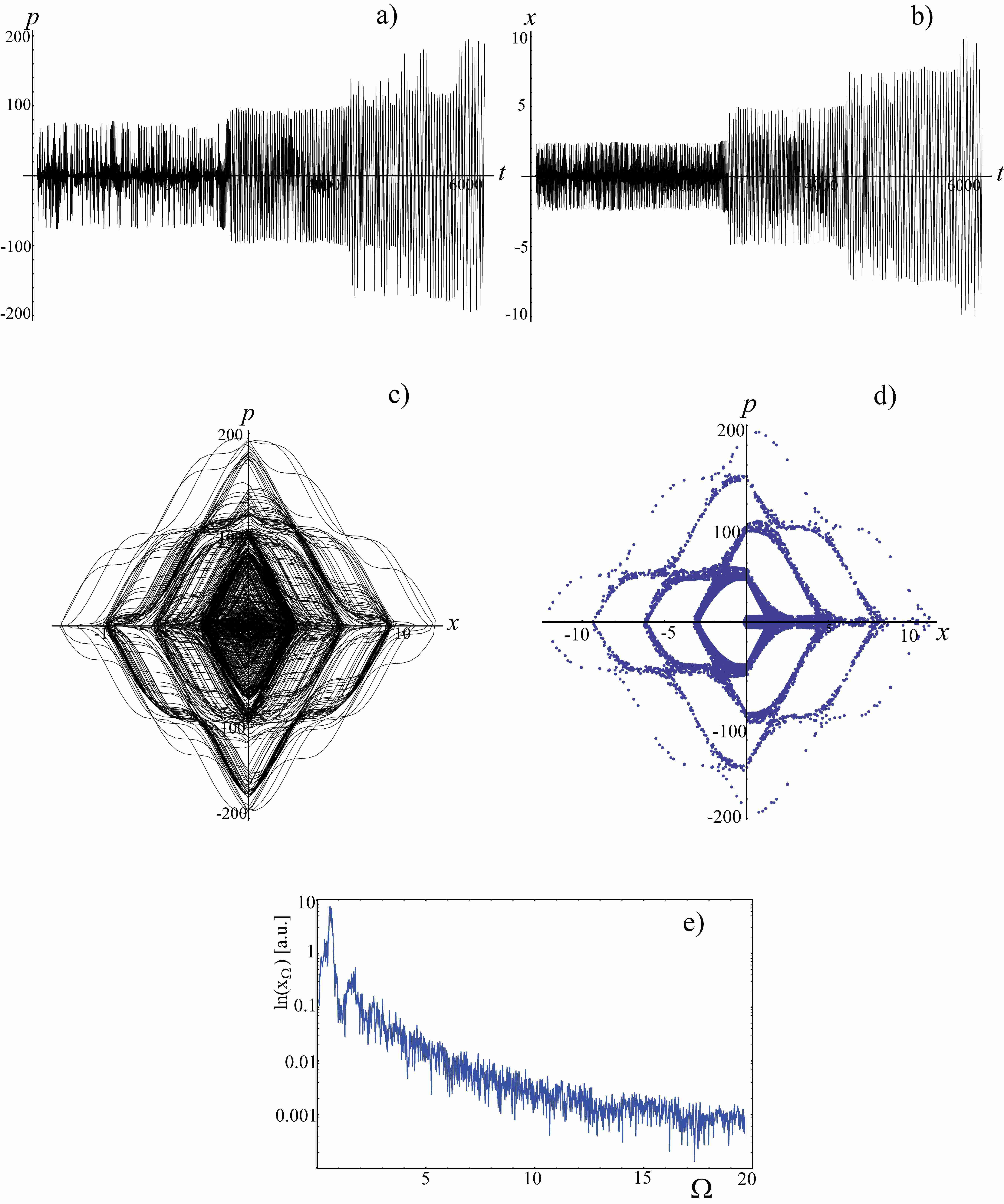}
  \end{center}
  \caption{Driven oscillations near the stochastic regime for 
  $\varepsilon_p=15.708$ and $a=0.62 (\pi \varepsilon_p/2)=15.298$. 
  a)  The momentum $p$ v.s. time. 
  b) The coordinate $x$ v.s. time. 
  c) The phase plane $(x,p)$.
    d) The Poincar\'{e} section showing the particle positions 
  in the phase plane $(x,p)$ at discrete times, with the time step 
equal to the period of the driving force, $2 \pi$.
 e) Fourier spectrum of the coordinate $x$. The dependence 
  of $\ln {x_{\Omega}}$ on the frequency  $\Omega$ is shown.}
     \label{Fig6}
\end{figure}

 Under the conditions corresponding to the parameters $\varepsilon_p$ and $a$ 
 above the threshold of the stochastic regime, 
 $a=0.62 (\pi \varepsilon_p/2)$ the features of the driven 
 nonlinear oscillators are shown in Fig. \ref{Fig6}. 
 As one can see, at the initial stage for the time interval $t\in[0,60]$ 
 the oscillation amplitude is significantly smaller than 
 the driver amplitude (see also Fig. \ref{Fig7} below). Then
 it rises up abruptly. The maximum momentum oscillation amplitude as seen in 
 Figs. \ref{Fig6} (a) and (b) reaches a level approximately 
 10 times higher than the driver amplitude. 
 The time dependence 
 of the momentum,  Fig. \ref{Fig6} (a), and coordinate, Fig. \ref{Fig6} (b), 
 demonstrate the intermittency of the irregular behaviour. 
 This clearly demonstrates that the intervals of high amplitude, 
 low frequency oscillations with relatively large 
trajectory excursion from the $x=0$ plane randomly alternate 
with low amplitude, high frequency motion when 
the trajectory is confined in the vicinity of $x=0$ plane. 
The trajectory is localized within the domain 
of the phase plane shown in Fig. \ref{Fig6} (c). 
The trajectory tightly fills this domain demonstrating the ergodic property 
of the system under consideration. The domain size increases as time grows,
 i. e. the particle momentum is gradually growing.
In Fig. \ref{Fig6} (d) we plot the Poincar\'{e} section showing the particle positions 
  in the phase plane $(x,p)$ at discrete times, with the time step 
equal to the period of the driving force, $2 \pi$.
It has the form of a finite thickness web (it is also known 
as the Arnold web \cite{AWEB1, AWEB2}), 
which is broadened due to the stochastic nature of the particle motion.
The property of the particle to migrate in the phase plane along the web 
 is typical for the minimal chaos regimes \cite{Chernikov}. In Fig. \ref{Fig6} (e) 
 the Fourier spectrum of the coordinate $x$ is presented. Here the dependence 
  of $\ln {x_{\Omega}}$ on the frequency  $\Omega$ shows that the spectrum is continuous, 
which is one of the indications of the chaos regime \cite{SINAI}.   

The process of the particle energization in the initial stage is shown in  Fig. \ref{Fig7}. The intervals 
of regular, high frequency oscillations alternate with relatively short periods of nonadiabatic dependence 
of the momentum and coordinate on time. Around $t=\pi, \, 2\pi, 3\pi, \,$ and 
so on, the trajectory is ``kicked'' 
by the driver field resulting in the momentum Fig. \ref{Fig7} (a) and coordinate 
\ref{Fig7} (b) amplitude increase.  
In the nonadiabatic phase the trajectory excursion along the $x$-coordinate is substantially larger 
than the oscillation amplitude in the adiabatic phase. 
Having been ``kicked'' several times the trajectory leaves the vicinity of the equilibrium 
plane $x=0$.

To compare the phase patterns for the parameters near and above the stochasticity 
threshold in Fig. \ref{Fig8}
 we plot the Poincar\'{e} sections showing the particle positions in the phase plane $(\phi,p)$ 
  at discrete times  with the time step 
corresponding to the crossing of the $x=0$ plane by the trajectory.
In the case presented in Fig. \ref{Fig8} (a) for 
  $\varepsilon_p=15.708$ and $a=0.62 (\pi \varepsilon_p/2)=15.298$ as in Fig. \ref{Fig6}, 
  the driven oscillator 
is near the threshold of the stochastic regime. 
The phase plane shows the overlapping relatively narrow resonances.
  Slightly increasing the driver amplitude, with $\varepsilon_p=15.708$ 
  and $a=0.8 (\pi \varepsilon_p/2)=19.74$, we obtain the oscillations 
  in the developed stochastic regime as seen 
  in in Fig. \ref{Fig8} (b). We see the overlapping of the broad stochastic layers
 similar to the pattern which is presented in Fig. \ref{Fig1} (b).

\subsection{Time dependence of the average kinetic energy}

In the limit of large driver amplitude, $a\gg \varepsilon_p$, the nonlinear oscillations are regular. 
In order to 
determine the parameters of domains where the oscillations are stochastic and where they are regular, 
we analyze the 
time average of the kinetic energy,
\begin{equation}
w(t_m)=\frac{1}{(1+a^2)^{1/2}\, t_m}\int_0^{t_m}\left[ (1+p(t)^2)^{1/2}-1\right]dt.
\label{eq:w-t}
 \end{equation}
Here $t_m$ is the averaging time. 
The electron kinetic energy $\overline{{\cal E}_e} $ is normalized by $m_e c^2(1+a^2)^{1/2}$. 
In the 
case of particle oscillations in the driver field only, i. e.  $\varepsilon_p=0$ which is similar to 
the mean electron energy introduced in Ref. \cite{TNSAW}, we have
\begin{equation}
w_0(t_m)=\frac{2}{\pi (1+a^2)^{1/2}}
\left[
\frac{1}{t_m}E\left(t_m;-{a^2}\right)-1
\right]
\label{eq:w0-t}
 \end{equation}
with the incomplete elliptic integral $E(x;k)$. In the limit $t_m \to \infty$
it yields for the time average of the normalized kinetic energy, 
\begin{equation}
w_0\approx \frac{2 E(-a^2)}{\pi (1+a^2)^{1/2}}.
 \end{equation}
For $a \gg 1$ we have $w_0\approx 2/\pi$. 

\begin{figure}[]
  \begin{center}
    \includegraphics[height=2.5cm, width=8cm]{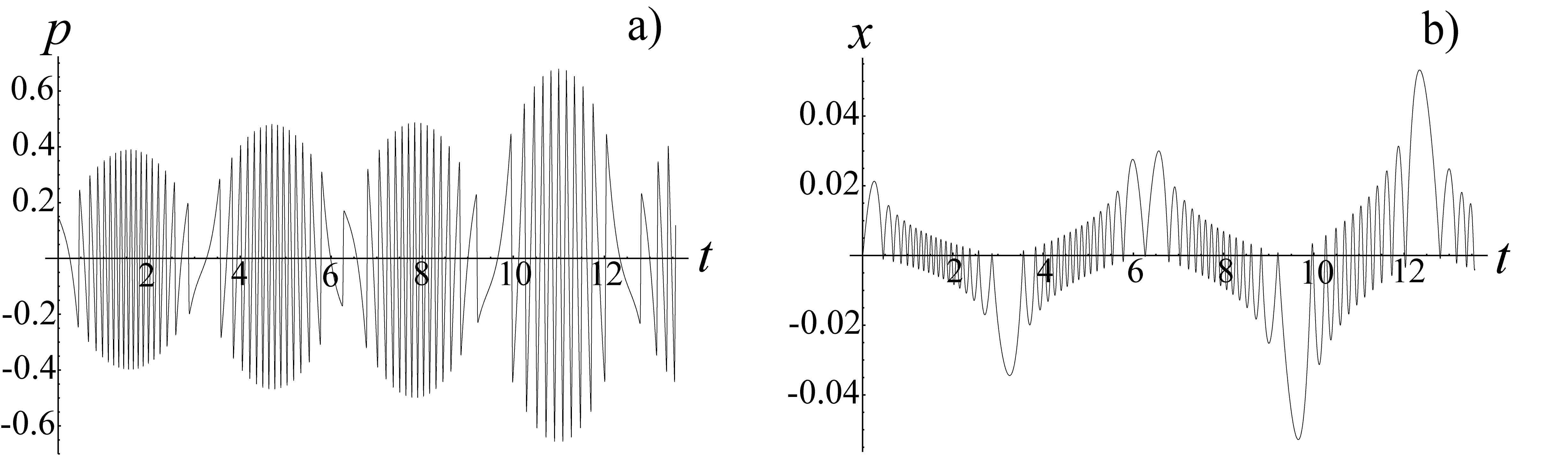}
  \end{center}
  \caption{Initial stage of driven oscillations in the near resonance conditions 
  for
  $\varepsilon_p=15.708$ and $a=0.62 (\pi \varepsilon_p/2)=15.298$. 
   at $0<t<4 \pi$. 
  a) The momentum $p$ v.s. time. 
  b) The coordinate $x$ v.s. time. 
  }   \label{Fig7}
\end{figure}

\begin{figure}[]
  \begin{center}
    \includegraphics[height=5.4cm, width=7.6cm]{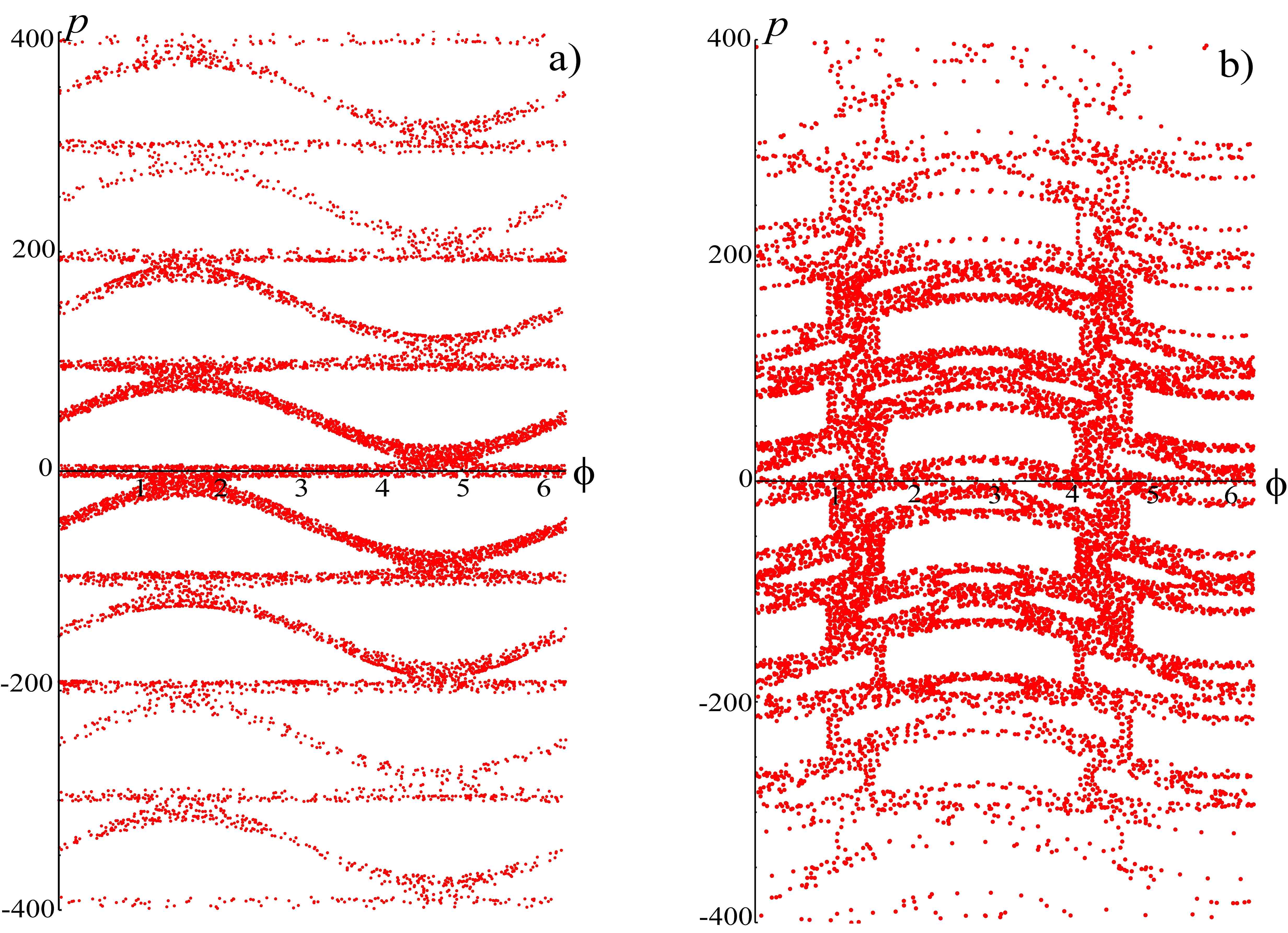}
  \end{center}
  \caption{  
  The Poincar\'{e} section showing the particle positions in the phase plane $(\phi,p)$ 
  at discrete times  with the time step 
corresponding to the crossing of the $x=0$ plane by the trajectory:
a) near threshold of the stochastic regime for 
  $\varepsilon_p=15.708$ and $a=0.62 (\pi \varepsilon_p/2)=15.298$;
  b) in the developed stochastic regime for 
  $\varepsilon_p=15.708$ and $a=0.8 (\pi \varepsilon_p/2)=19.74$.
  }
     \label{Fig8}
\end{figure}
\begin{figure}[]
  \begin{center}
    \includegraphics[height=2.7cm, width=8.1cm]{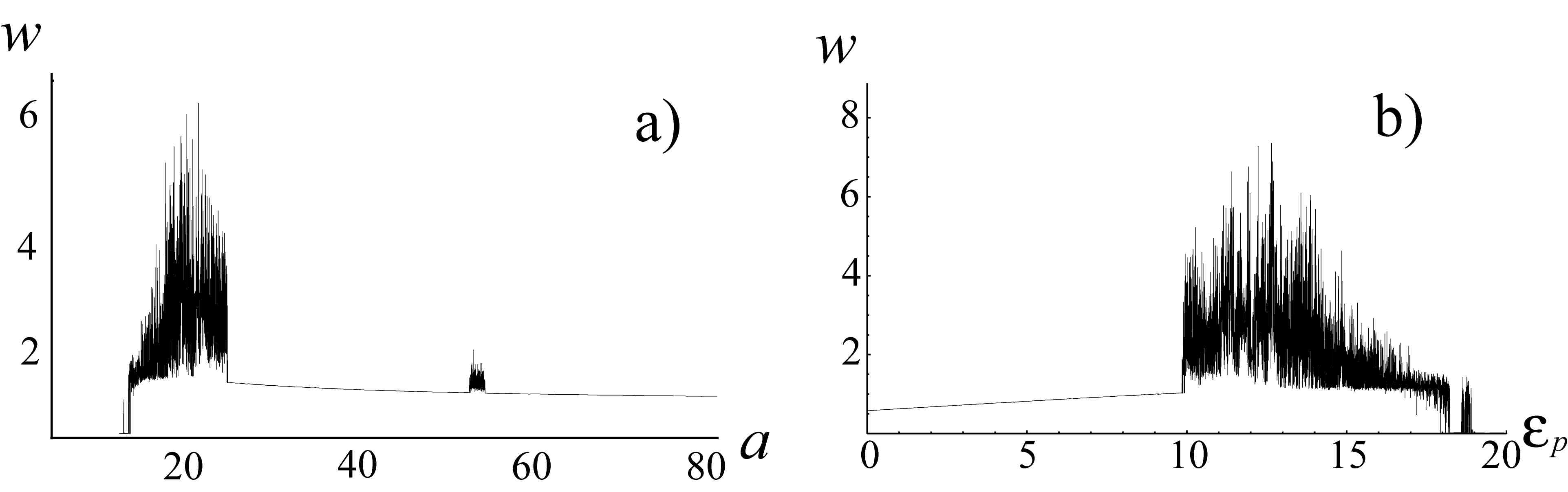}
  \end{center}
  \caption{Average normalized kinetic energy as a function of the driver amplitude $a$ for the 
  given parameter $\varepsilon_p=15.708$, (a), and as a function of the parameter $\varepsilon_p$ 
  for given amplitude $a=15.298$, (b). The averaging time is $t_m=100\times 2 \pi$.
  }   \label{Fig9}
\end{figure}
\begin{figure}[]
  \begin{center}
    \includegraphics[height=2.7cm, width=4.4cm]{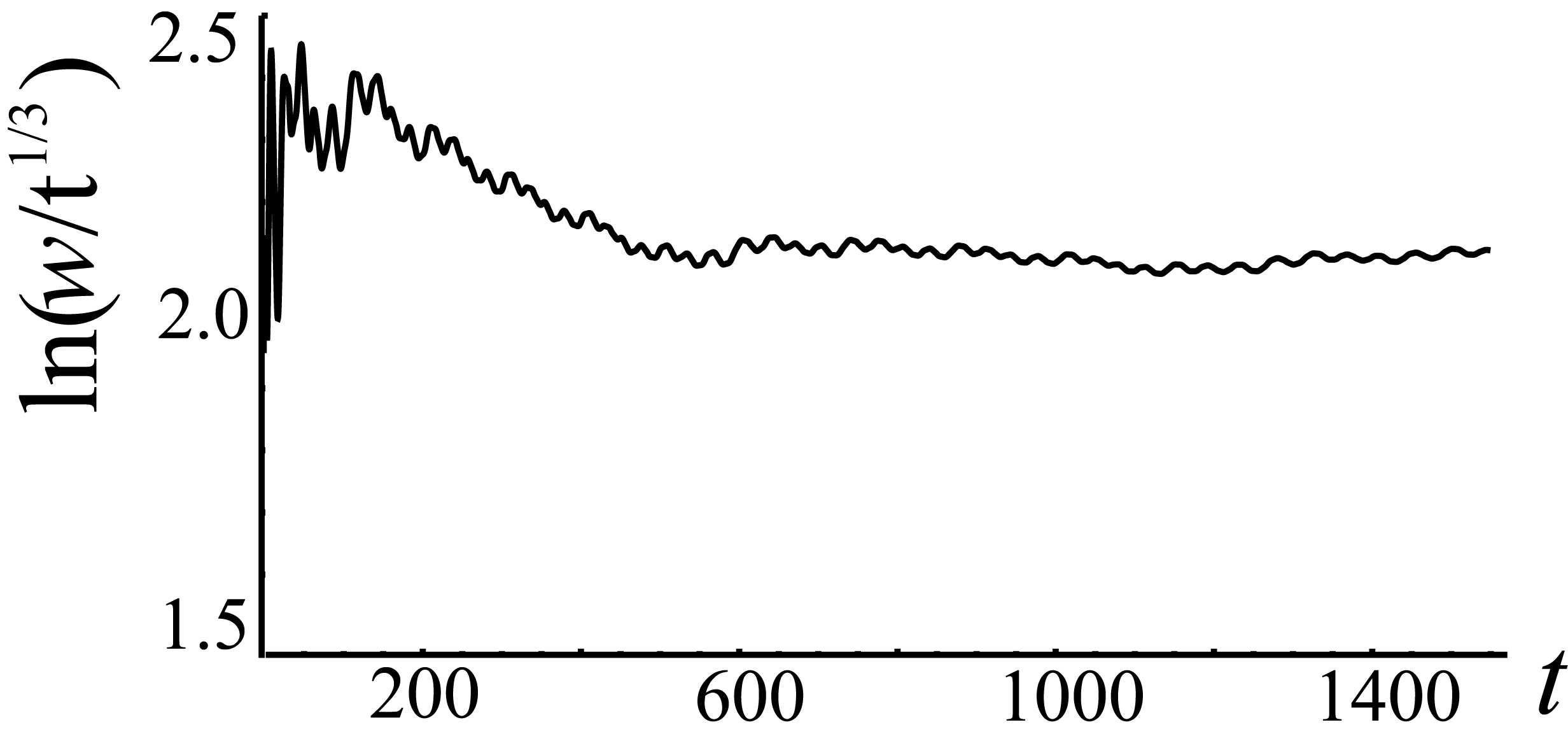}
  \end{center}
  \caption{Dependence on time of the average normalized kinetic energy logarithms 
  for the driver amplitude $a=15.708$  
  parameter $\varepsilon_p=11.781$.  }   
  \label{Fig10}
\end{figure}

In Fig. \ref{Fig9} we plot the parametric dependences of the time averaged normalized kinetic energy, $w$.
Fig. \ref{Fig9} (a) shows the function $w(a)$ for fixed parameter $\varepsilon_p$. 
In Fig. \ref{Fig9} (b) we 
plot the function $w(\varepsilon_p)$ for fixed driver amplitude $a$. 
The dependences of the average normalized kinetic 
energy on the parameters $a$ and $\varepsilon_p$ clearly demonstrate 
the stochastic regions above the threshold $a=0.62 (\pi \varepsilon_p/2)$ with the width of the order of 
$\delta a\approx 10$ and $\delta \varepsilon_p\approx 10$. 
The analysis of the time dependence of the kinetic energy shows (see Fig. \ref{Fig10}) that its growth 
approximately corresponds to the dependence $\propto t^{\kappa}$. Within a relatively short time interval 
the index $\kappa$ is close to $1/2$ as predicted by Eq. (\ref{eq:p-treg}) 
although there is a significant uncertainty. Over a relatively long 
time interval $t\in [100,1500]$ the averaged energy grows proportionally to $t^{1/3}$ 
according to Eq. (\ref{eq:p-tstoch}) as seen in Fig. \ref{Fig10}, 
where the time dependence of $\ln(w/t^{1/3})$ is plotted.

\begin{figure}[]
  \begin{center}
    \includegraphics[height=6.3cm, width=8.1cm]{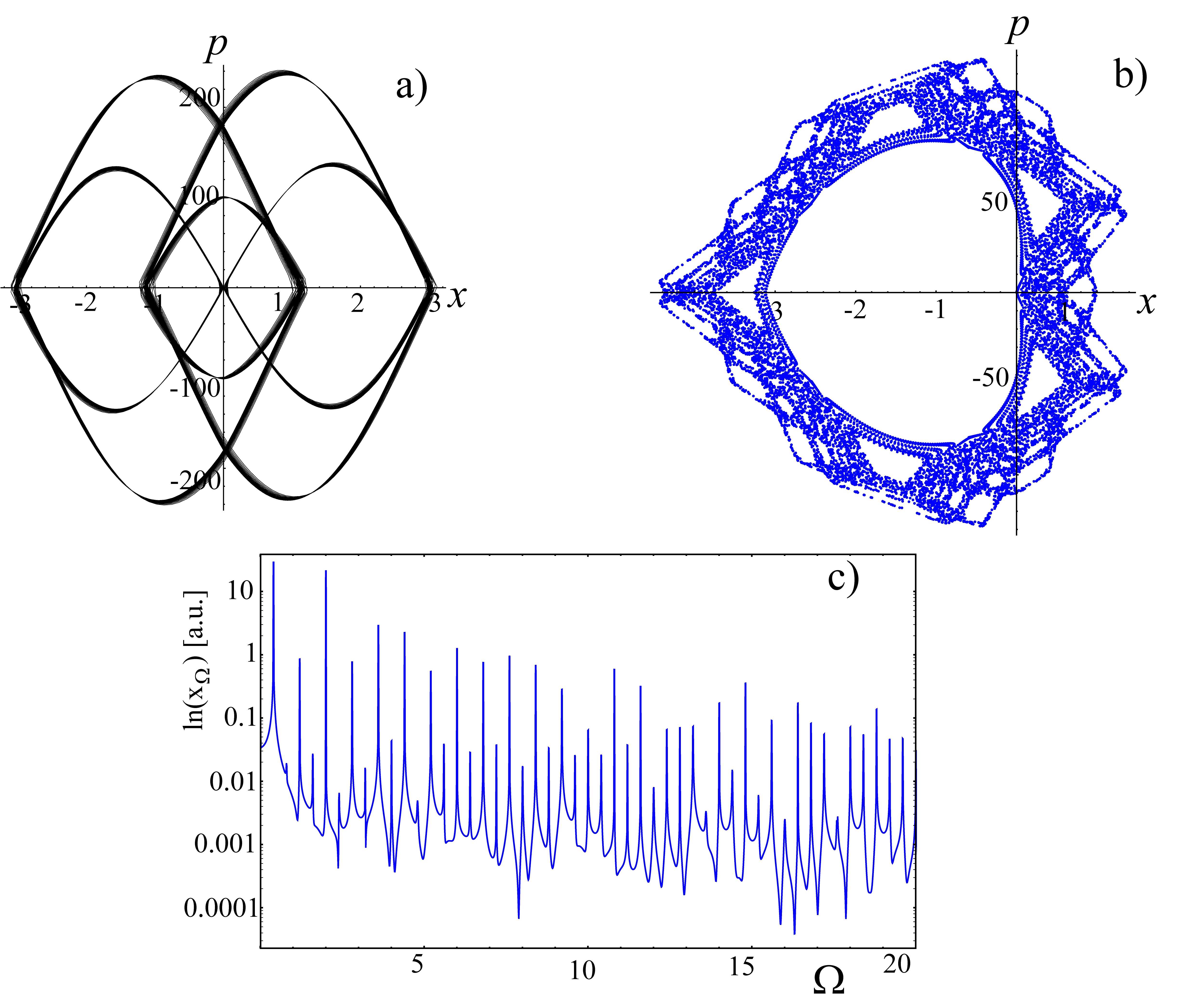}
  \end{center}
  \caption{Driven oscillations near stochastic regime for 
  $\varepsilon_p=15.708$ and $a=9.986\times 0.62 (\pi \varepsilon_p/2)=152.765$. 
  a) The phase plane $(x,p)$.
  b) The Poincar\'{e} section showing the particle positions 
  in the phase plane $(x,p)$ at discrete times, with the time step 
equal to the period of the driving force, $2 \pi$.
   c) Fourier spectrum of the coordinate $x$. It is shown the dependence 
  of $\ln {x(\Omega)}$ on the frequency  $\Omega$.
  }   \label{Fig11}
\end{figure}

\begin{figure}[]
  \begin{center}
    \includegraphics[height=6.3cm, width=8.1cm]{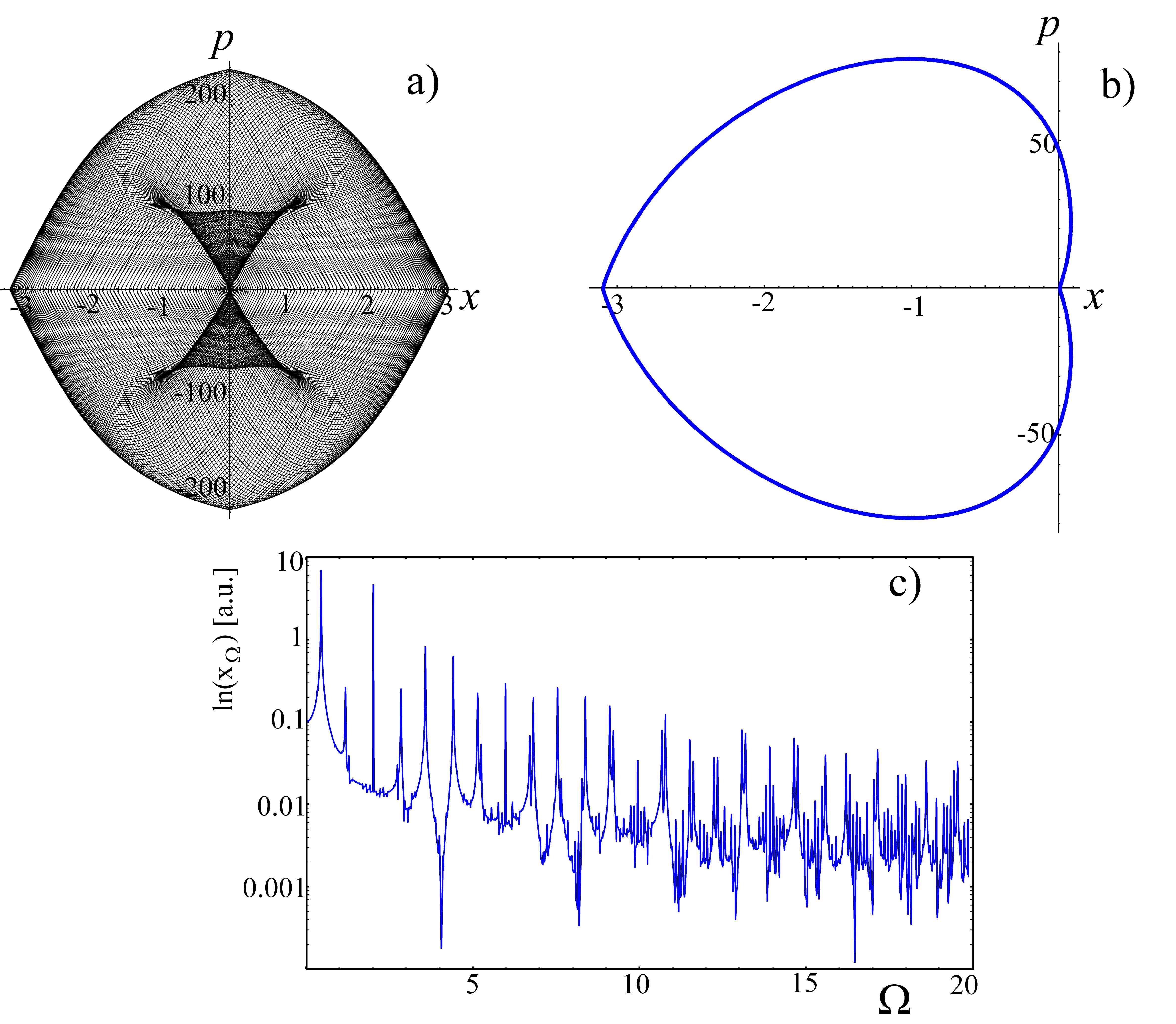}
  \end{center}
  \caption{Driven oscillations in regular regime for 
  $\varepsilon_p=15.708$ and $a=9\times 0.62 (\pi \varepsilon_p/2)=137.681$. 
  a) The phase plane $(x,p)$.
  b) The Poincar\'{e} section showing the particle positions 
  in the phase plane $(x,p)$ at discrete times, with the time step 
equal to the period of the driving force, $2 \pi$.
  c) Fourier spectrum of the coordinate $x$. The dependence 
  of $\ln {x(\Omega)}$ on the frequency  $\Omega$ is shown.
  }   \label{Fig12} 
\end{figure}

In Fig. \ref{Fig9} we can see another stochastic region in the interval $a \in [50,60]$. 
In Figs. \ref{Fig11} and \ref{Fig12} we present 
two examples of regular dynamics described by the solutions of Eqs. (\ref{eq:dotp},\ref{eq:dotx}) 
for the parameters beyond the stochastic region, i. e. $a\ll \varepsilon_p$.

In Fig. \ref{Fig11} we plot the phase plane $(x,p)$, 
Poincar\'{e} section and the Fourier spectrum of the coordinate $x$
 for the driver amplitude and the parameter 
$\varepsilon_p$ equal to 15.708 and 
$a=9.986\times 0.62 (\pi \varepsilon_p/2)$, respectively. 
The trajectory in the phase plane is localized in narrow 
width stripes. The Poincar\'{e} section shows 
a stochastic layer with islands around the periodic trajectory. However, the Fourier spectrum 
is comprised of high order harmonics, which is typical for regular nonlinear oscillations. 

Slightly changing the driver amplitude to $a=9\times 0.62 (\pi \varepsilon_p/2)$ with the same parameter
$\varepsilon_p$ as in Fig. \ref{Fig11} results in ergodic motion,  
when the trajectory completely 
fills the confinement domain in the phase plane (see \ref{Fig12} (a).  
According  to the Poincar\'{e} section in Fig. \ref{Fig12} (b) and the 
Fourier transform, Fig. \ref{Fig12} (c), consisting of high order harmonics,
 the nonlinear oscillations are regular 
with no stochastic layer.

Note that the stochastic-like regimes with developed 
stochastic layers seen in Figs. \ref{Fig11} (b) 
are not ergodic, i.e. the trajectory being confined within the finite 
size domain in the phase plane forms 
finite width stripes. 
It does not fill the domain. Contrary to this, there is an example when 
the regular trajectory fills the confinement 
domain as is seen in Fig. \ref{Fig12} (a).   

 \section{4. Conclusion}
 \label{Sec4}
 
In conclusion, we have demonstrated that a driven oscillator with a step-like nonlinearity 
can exhibit regular and chaotic oscillation regimes.    The step-like nonlinearity implies 
a maximal value of the restoring force. The threshold of the stochastic 
oscillation onset corresponds  to the 
condition of a strong driver  compared with the restoring force.
It can be written 
as a relationship between the normalized electromagnetic field amplitude and the normalized surface density:
 $a>\varepsilon_p$.  This inequality is equivalent to the criterion of thin slab relativistic transparency 
\cite{VAV, ThinFoil1, MIRR, REVFP}. It also determines the parameters optimal for ion acceleration by the 
laser light radiation pressure \cite{MIRR, REVFP, RPDA, SSB12}. The model used above describes 
the anomalous electron heating due to the electron recirculation around thin foil target. 
We note that the electron recirculation during the ion acceleration by the light radiation pressure 
has been seen in theoretical work presented in Refs. \cite{ThinFoil1} and \cite{Unlim1, Unlim2}.
The electron heating can be enhanced for the tightly focused laser pulse \cite{Dollar} and for not 
perfectly aligned laser-target configurations \cite{Lezhnin}.

Under the chaos conditions the oscillation amplitude grows proportionally with time to the one third power, 
which is an indication of the trajectory diffusion in phase space previously discussed 
in Ref. \cite{ThinFoil0}. 
This result has been obtained by analyzing the Poincar\'{e} mapping, by solving the 
KFPE diffusion equation, and by 
numerical integration of an ensemble of nonlinear oscillator equations. 

The obtained electron energy scaling given by Eq. (\ref{eq:p-tstochEn}), at the stochasticity threshold  
$a\approx\varepsilon_p$ in the limit $a\gg 1$ can be approximated by

\begin{equation}
\frac{\overline{{\cal E}_e}}{ m_e c^2} \approx 1.11 \times
\left(I\,\left[\frac{\rm EW}{{\rm cm}^2}\right]\right)^{1/2}
\lambda\,[\mu{\rm m}]
(\tau_{las}\,[{\rm fs}])^{1/3}, 
 \end{equation}
where $\tau_{las}$ is the laser pulse duration. This scaling is favorable for achieving 
high energy ion beams generated under the conditions when the TNSA acceleration mechanizm is realized in the 
laser interaction with thin foil targets.  This shows substantially higher electron temperature than 
the scaling found in Ref. \cite{TNSAW}. This is due to the electron recirculation around a thin foil target 
as has been found in PIC simulations 
in Refs.  \cite{RECIRC, LMM-2015} (see also Refs. \cite{Taguchi, Aref}) 
resulting in anomalous electron heating.

 \section*{Acknowledgments}
 
This work was supported by a Scientific Research (C) No. 25420911 
commissioned by MEXT and partially supported by NEXT Program of JSPS.
SVB is grateful to Prof. S. I. Krasheninnikov for discussions 
and to Prof. F. Pegoraro for useful comments.
AY appreciate fruitful discussions with Prof. S. Fujioka and H. Azechi of ILE.
TZhE acknowledges the support from JSPS (Grant No. 25390135).

\end{document}